# Energetic Particle Increases Associated with Stream Interaction Regions

C. M. S. Cohen[1], E. R. Christian[2], A. C. Cummings[1], A. J. Davis[1], M. I. Desai[3], J. Giacalone[4], M. E. Hill[5], C. J. Joyce[6], A. W. Labrador[1], R. A. Leske[1], W. H. Matthaeus[7], D. J. McComas[6], R. L. McNutt, Jr.[5], R. A. Mewaldt[1], D. G. Mitchell[5], J. S. Rankin[6], E. C. Roelof[5], N. A. Schwadron[8], E. C. Stone[1], J. R. Szalay[6], M. E. Wiedenbeck[9], R. C. Allen[5], G. C. Ho[5], L. K. Jian[2], D. Lario[2], D. Odstrcil[2,10], S. D. Bale[11], S. T. Badman[11], M. Pulupa[12], R. J. MacDowall[13], J. C. Kasper[14], A. W. Case[15], K. E. Korreck[15], D. E. Larson[16], Roberto Livi[16], M. L. Stevens[15], and Phyllis Whittlesey[16]
[1] California Institute of Technology, Pasadena, CA 91125, USA; cohen@srl.caltech.edu
[2] NASA/Goddard Space Flight Center, Greenbelt, MD 20771, USA
[3] University of Texas at San Antonio, San Antonio, TX 78249, USA
[4] University of Arizona, Tucson, AZ 85721, USA
[5] Johns Hopkins University Applied Physics Laboratory, Laurel, MD 20723, USA
[6] Department of Astrophysical Sciences, Princeton University, Princeton, NJ 08544, USA
[7] University of Delaware, Newark, DE 19716, USA
[8] University of New Hampshire, Durham, NH 03824, USA
[9] Jet Propulsion Laboratory, California Institute of Technology, Pasadena, CA 91109, USA
[10] George Mason University, Fairfax, VA 22030, USA
[11] Physics Department, University of California at Berkeley, Berkeley, CA 94720, USA
[12] Space Sciences Laboratory, University of California at Berkeley, Berkeley, CA 94720, USA
[13] Solar System Exploration Division, NASA/Goddard Space Flight Center, Greenbelt, MD 20771, USA
[14] University of Michigan, Ann Arbor, MI 48109, USA
[15] Smithsonian Astrophysics Observatory, Cambridge, MA 02138, USA
[16] University of California at Berkeley, Berkeley, CA 94720, USA
Received 2019 September 13; revised 2019 October 4; accepted 2019 October 7; published 2020 February 3

## Abstract

The *Parker Solar Probe* was launched on 2018 August 12 and completed its second orbit on 2019 June 19 with perihelion of 35.7 solar radii. During this time, the Energetic Particle Instrument-Hi (EPI-Hi, one of the two energetic particle instruments comprising the Integrated Science Investigation of the Sun, IS☉IS) measured seven proton intensity increases associated with stream interaction regions (SIRs), two of which appear to be occurring in the same region corotating with the Sun. The events are relatively weak, with observed proton spectra extending to only a few MeV and lasting for a few days. The proton spectra are best characterized by power laws with indices ranging from $-4.3$ to $-6.5$, generally softer than events associated with SIRs observed at 1 au and beyond. Helium spectra were also obtained with similar indices, allowing He/H abundance ratios to be calculated for each event. We find values of 0.016–0.031, which are consistent with ratios obtained previously for corotating interaction region events with fast solar wind $\leqslant 600\,\mathrm{km\,s^{-1}}$. Using the observed solar wind data combined with solar wind simulations, we study the solar wind structures associated with these events and identify additional spacecraft near 1 au appropriately positioned to observe the same structures after some corotation. Examination of the energetic particle observations from these spacecraft yields two events that may correspond to the energetic particle increases seen by EPI-Hi earlier.

*Unified Astronomy Thesaurus concepts:* Solar energetic particles (1491); Solar activity (1475); Fast solar wind (1872)

## 1. Introduction

Energetic particle increases associated with stream interaction regions (SIRs) have been studied for decades (Bryant et al. 1965; Wilcox & Ness 1965; McDonald & Desai 1971; Barnes & Simpson 1976; Zwickl & Roelof 1981; Richardson et al. 1993). These regions are typically formed when a stream of faster solar wind overtakes slower solar wind; when the speed differential is substantial enough, a shock or significant compression region can form capable of accelerating particles to MeV nuc$^{-1}$ energies (see a review by Richardson 2004). Often the high-speed stream originates in a solar coronal hole that may persist for multiple rotations causing the SIR to reoccur or corotate with the Sun; in these instances the structure is referred to as a corotating interaction region (CIR).

Due to the expansion of the CIR with an increasing distance from the Sun, the edges of the compression region can steepen into a pair of shocks; the one propagating into the slow solar wind is referred to as the "forward" shock, while the one propagating sunward in the solar wind frame into the fast stream is the "reverse" shock. Although either shock can accelerate particles, it is most often the reverse shock that is associated with energetic particle increases seen at 1 au (Fisk & Lee 1980; Giacalone & Jokipii 1997; Richardson 2004).

Although solar energetic particle (SEP) events also involve shock acceleration, several of the properties of the energetic particle population are distinctly different in SEP and CIR events. Generally, the particle spectra in CIRs are significantly softer than those in SEP events at energies of a few MeV nuc$^{-1}$; e.g., power-law indices are typically $-4$ in CIR events and approximately $-3$ in SEP events (although there is a fair amount of variability in SEP spectra). The particle intensity enhancements in CIR events have more gradual onsets often with no velocity dispersion and are fairly isotropic in pitch angle, unlike the abrupt, dispersive, anisotropic populations in SEP events. Finally, the elemental composition is different in the two populations, particularly for He/H, which is typically a factor of ∼3 lower in SEP events (Richardson 2004 and references therein).

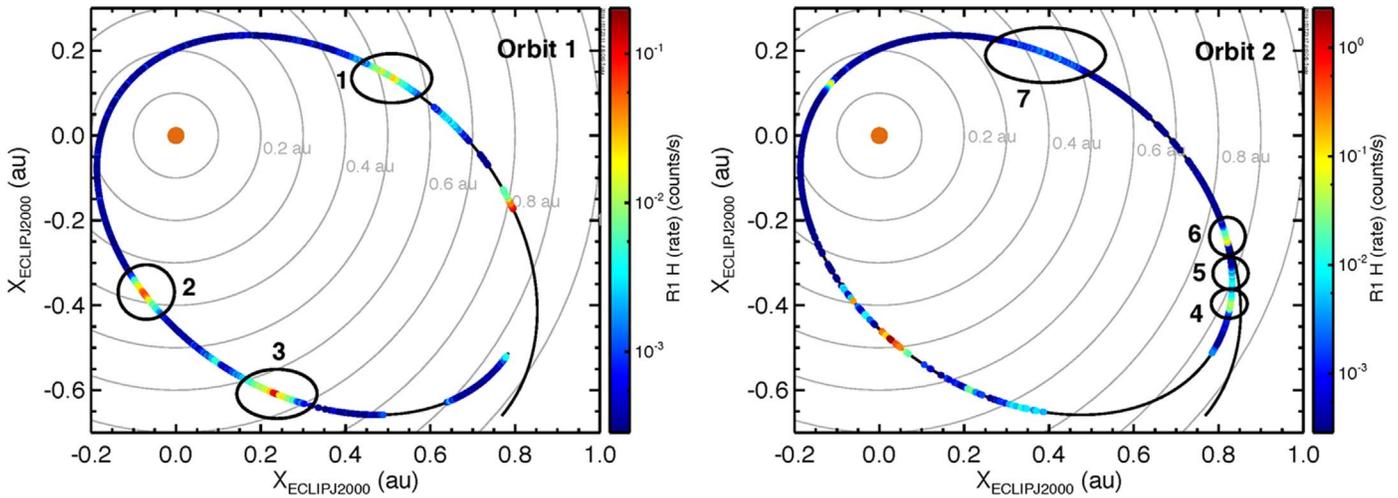

**Figure 1.** First two orbits of *PSP* (2018 September 25–2019 January 20, left, and 2019 January 20–2019 June 19, right). The color intensity shows ∼1–2 MeV proton counting rate measured by LETA. The seven events of this study are identified by ovals and numbers (additional events are discussed in McComas et al. 2019; Leske et al. 2020; and Wiedenbeck et al. 2020). The location of the Sun is given by the orange circle and radial distances are marked by the gray concentric circles.

Most of our knowledge of CIR events is from observations near 1 au and beyond. However, the Helios spacecraft observed several CIR events, including five inside 0.6 au (Van Hollebeke et al. 1978). The recently launched *Parker Solar Probe* (*PSP*) provides another opportunity to study such events well inside 1 au and to compare the energetic particle population characteristics to those observed at larger distances. The first two orbits of *PSP* yielded seven periods of enhanced intensities of energetic particles that are likely related to SIRs/CIRs (McComas et al. 2019). Here we present more detailed analyses of these events to reveal some characteristics of protons and helium at energies of a few MeV nuc$^{-1}$.

## 2. Instrumentation

*PSP* (Fox et al. 2016) was launched on 2018 August 12 and placed into an orbit that, with the help of seven Venus flybys, will carry the spacecraft to within 10 solar radii ($R_S$) of the Sun by the end of 2024. Even in the first orbit (with perihelion of 35.7 $R_S$), *PSP* explored regions closer to the Sun than any spacecraft has before. The spacecraft carries four science packages, designed to measure the solar wind plasma and electromagnetic fields, and energetic particles, and to image the solar corona and inner heliosphere: the Solar Wind Electrons Alphas and Protons Investigation (SWEAP; Kasper et al. 2016), the Electromagnetic Fields Investigation (FIELDS; Bale et al. 2016), the Integrated Science Investigation of the Sun (IS☉IS; McComas et al. 2016), and the Wide Field Imager for Solar Probe Plus (WISPR; Vourlidas et al. 2016).

Two energetic particle instruments comprise IS☉IS, including one measuring the lower energies (the Energetic Particle Instrument-Lo, EPI-Lo) and one focusing on the higher energies (EPI-Hi); a complete description of these instruments is given in McComas et al. (2016), Hill et al. (2017), and Wiedenbeck et al. (2017). Once EPI-Lo and EPI-Hi are fully intercalibrated (see Joyce et al. 2020), it will be possible to combine the measurements to provide energy spectra spanning nearly four orders of magnitude. However, this has yet to be achieved, particularly for protons where the EPI-Lo instrument response is complicated. Thus, for this study, we have limited the analysis to EPI-Hi observations, and here we detail only the EPI-Hi capabilities.

The EPI-Hi instrument consists of two double-ended and one single-ended "telescopes" that are each formed of a stack of silicon detectors allowing particle identification via the standard dE/dx versus energy technique. Ions with energies above ∼10 MeV nuc$^{-1}$ are measured by the high-energy telescope (HETA and HETB for the sunward and anti-sunward directed ends, respectively); energies below ∼20 MeV nuc$^{-1}$ are covered by the low-energy telescopes, LET1 being the double-ended sensor and LET2 being the single-ended sensor. We refer to the two ends of LET1 as LETA (sunward facing) and LETB (anti-sunward) and the single aperture of LET2 as LETC. The center of the field of view of LETC is orthogonal to those of LETA and LETB, allowing the full range of particle pitch angles to be observed.

The events presented here were too weak to be measurable by HET, thus our analysis is further focused on LET observations. Each LET aperture has a geometry factor that ranges from 0.25 cm$^2$ sr at ∼1 MeV nuc$^{-1}$ to 0.63 cm$^2$ sr at ∼10 MeV nuc$^{-1}$. Both ends of LET1 have ∼12 $\mu$m detectors at the top of the stack, allowing particles with energies as low as 1 MeV nuc$^{-1}$ to be measured. In contrast, LETC has a thicker window protecting the entrance, which results in minimum energies of ∼2 MeV nuc$^{-1}$.

## 3. Observations

Seven particle increases observed by EPI-Hi were identified in the first two orbits of *PSP* as SIR events (Figure 1); other particle increases associated with SEP events are discussed in McComas et al. (2019), Wiedenbeck et al. (2020), and Leske et al. (2020). Three of the events occurred when *PSP* was inside of ∼0.6 au, which is substantially closer to the Sun than most studied SIR/CIR events; Helios reported five events at these distances (Van Hollebeke et al. 1978). The top panel of Figure 2 shows the energy and time variation of the seven events as observed by LETA. The events are relatively small with proton intensity increases evident up to only a few MeV

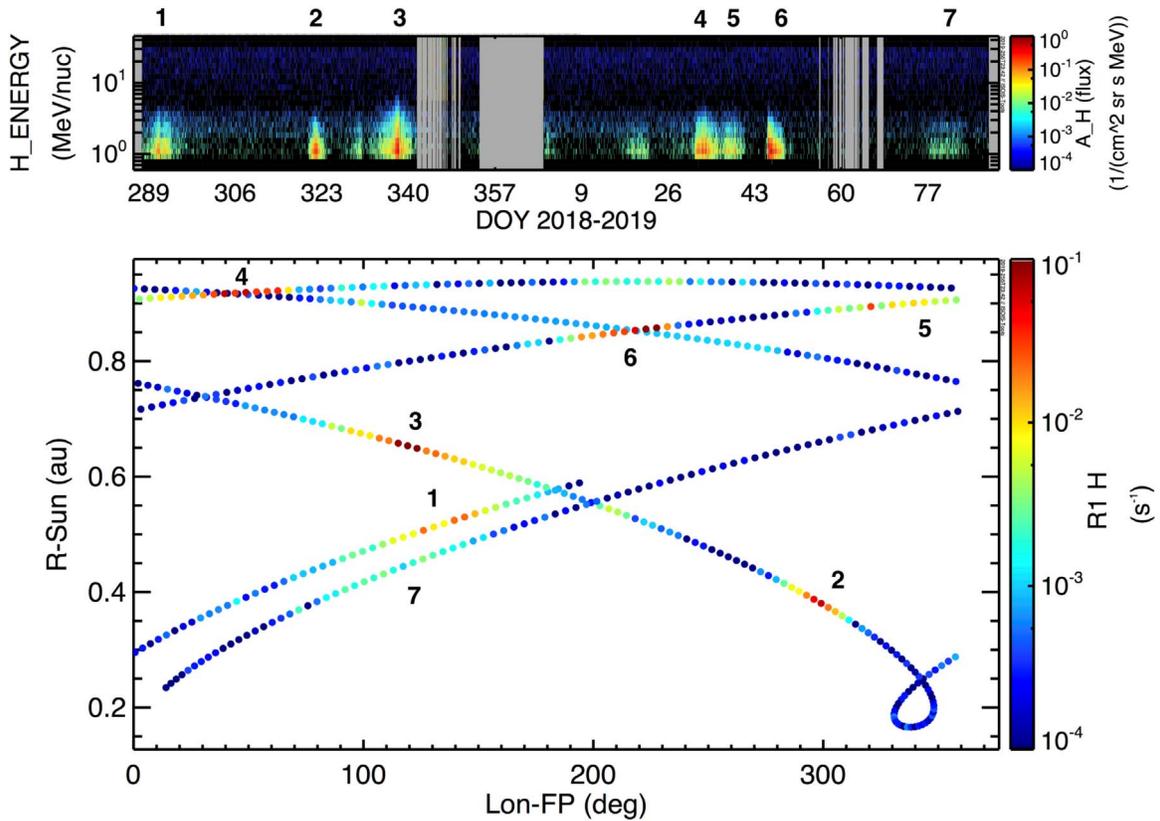

**Figure 2.** (Top panel) Proton intensity spectrogram from LETA showing the seven events (as numbered along the top of the panel) as a function of time and energy. Gray regions are data gaps. (Bottom panel) Counting rate of 1–2 MeV protons from LETA (the color bar) as a function of the distance from the Sun and the longitude of the spacecraft's magnetic footpoint (assuming a Parker spiral corresponding to 400 km s$^{-1}$). Increases corresponding to the seven events are labeled by number; time increases with the decreasing footpoint longitude along each trajectory segment.

but that last for several days (the specific time periods selected are given in Table 1).

The magnetic footpoint of the spacecraft was determined by calculating the Parker spiral connecting the Sun and *PSP*, assuming a solar wind speed of 400 km s$^{-1}$. The heliocentric distance of the spacecraft is plotted versus the calculated footpoint in Figure 2. The color intensity of the points corresponds to the ∼1–2 MeV LETA proton counting rate (in counts-sec$^{-1}$). Events 1 and 3 (2018 October 14 and 2018 November 28, respectively) are shown to have occurred at approximately the same solar footpoint longitude, suggesting these are corotating (i.e., CIR events; McComas et al. 2019). Event 7, observed 3.5 months later (2019 March 17), also occurs at the same solar longitude, yet there are no corresponding particle increases when *PSP* passed by the same longitude three times in the intervening interval. Thus, we consider this event unrelated to the previous two.

For each event, the proton intensities were integrated over the time periods given in Table 1 and corrected for background. The background spectrum was determined by integrating over many quiet days (as described in Leske et al. 2020) and is mostly due to galactic cosmic rays (and anomalous cosmic rays for He). The resulting spectra from the three EPI-Hi LET apertures are given in Figure 3 for each event. It should be noted that LETC has a thicker window at the entrance of the telescope than LETA and LETB, resulting in the LETC spectra starting at a higher energy than the LETA and LETB spectra. Although there are some variations, particularly at the higher energies, generally, the three spectra show good agreement indicating a fairly isotropic particle distribution. For further analyses, we have averaged the LETA and LETB spectra for each event.

Figure 4 compares the proton spectra of the seven events; a power law corresponding to $E^{-4.5}$ is also shown for reference. Most of the events are seen to be fairly well described by a power law with indices similar to −4.5. It should be noted that the dip in the spectra near 1.8 MeV is an instrumental effect; we have ignored this data point when fitting the spectra. The power-law indices resulting from fitting the spectra are given in Table 1 and range from −4.3 to −6.5.

Although the statistics are significantly poorer, He spectra were also obtained during each event. The LETA–LETB average spectra are shown in Figure 5. Generally, the helium spectra have similar spectral indices to the proton spectra. From these spectra, the He/H abundance ratio as a function of energy can be calculated and is shown in Figure 6. Due to the large uncertainties (primarily due to the poor He statistics), it is not possible to identify any energy dependence of the ratio in any of the events. The weighted mean of the He/H ratios in the lowest four energy bins (1.1–2.4 MeV nuc$^{-1}$) is given in Table 1 for each event.

Although EPI-Hi is capable of measuring ions heavier than He (see, e.g., observations during a small SEP event presented in Leske et al. 2020 and during a $^3$He-rich SEP event detailed in Wiedenbeck et al. 2020), sufficient fluxes have not been observed to date to properly calibrate the heavy ion response in the LET and HET telescopes. Regardless, based on nominal O/He abundance ratios observed in CIR

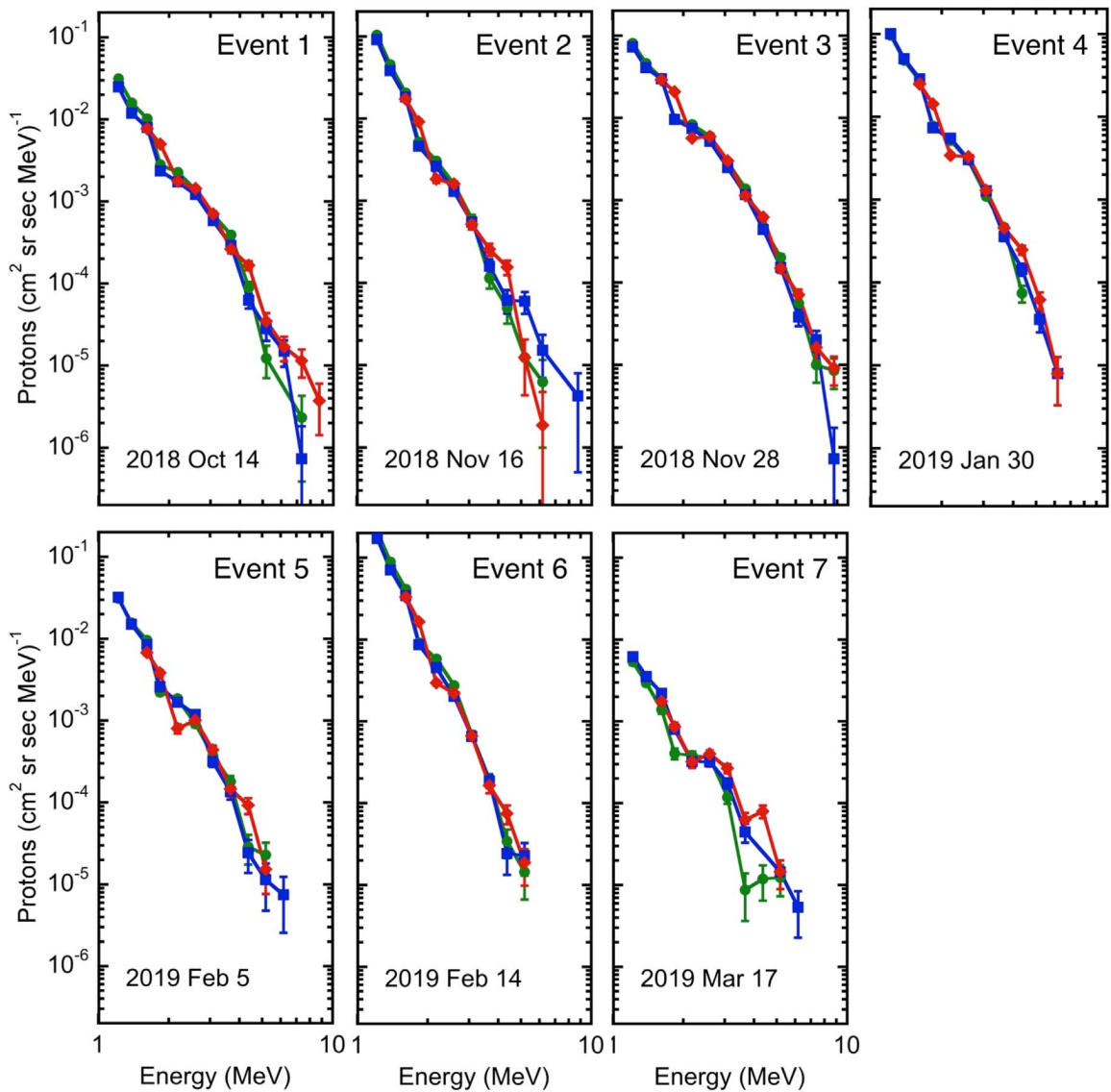

**Figure 3.** Time-integrated proton spectra for each event from LETA (green circles), LETB (blue squares), and LETC (red diamonds). The spectra have been corrected for background (primarily galactic cosmic rays) and uncertainties are statistical. The intensity dip near 1.8 MeV in all the LETA and LETB spectra (and near 2.2 MeV in LETC) is an instrumental effect.

**Table 1**
Selected Events

| Event Number | Time Period | $R$ (au) | H Spectral index | He/H (1.1–2.4 MeV nuc$^{-1}$) |
|---|---|---|---|---|
| 1 | 2018 Oct 14 15:00–2018 Oct 24 00:00 | 0.54 | −5.1 | 0.028 ± 0.001 |
| 2 | 2018 Nov 16 00:00–2018 Nov 20 00:00 | 0.38 | −5.6 | 0.031 ± 0.004 |
| 3 | 2018 Nov 28 00:00–2018 Dec 7 15:00 | 0.65 | −4.8 | 0.028 ± 0.003 |
| 4 | 2019 Jan 30 00:00–2019 Feb 5 12:00 | 0.92 | −5.1 | 0.028 ± 0.003 |
| 5 | 2019 Feb 5 12:00–2019 Feb 11 00:00 | 0.9 | −5.2 | 0.020 ± 0.005[a] |
| 6 | 2019 Feb 14 00:00–2019 Feb 19 00:00 | 0.85 | −6.5 | 0.031 ± 0.003 |
| 7 | 2019 Mar 17 00:00–2019 Mar 27 00:00 | 0.45 | −4.3 | 0.016 ± 0.007 |

**Note.**
[a] Averaged over 1.1–1.7 MeV nuc$^{-1}$.

events (Richardson 2004) and our measured He spectra, we find that over the entirety of the largest of our events, we would expect ∼0.3 counts of oxygen at 1–2 MeV nuc$^{-1}$. Even the higher O/He abundance ratio recently determined by Reames (2018) would result in only ∼0.5 oxygen counts. Thus, we are unable to comment on event composition beyond He/H, and it is likely that any oxygen observed by LET during these events is primarily anomalous cosmic rays.

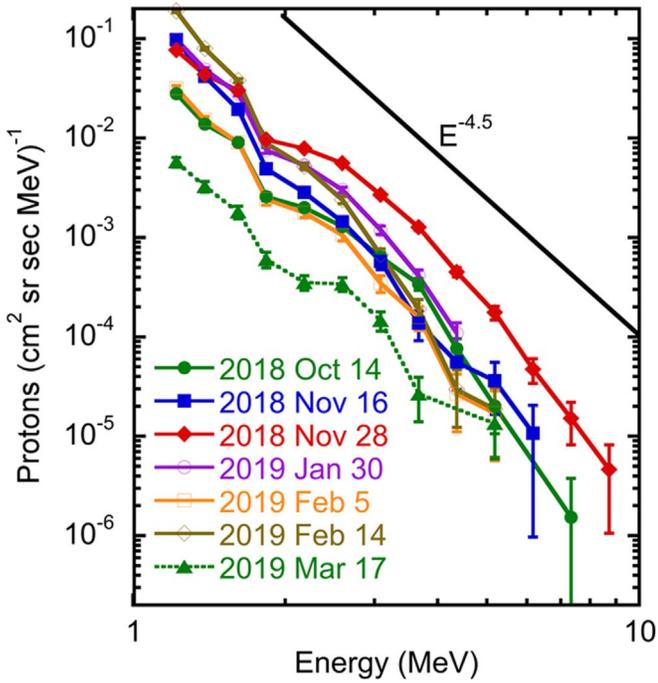

**Figure 4.** Time-integrated proton spectra for all seven events. These spectra are the average of the LETA and LETB spectra shown in the previous figure. A line corresponding to an $E^{-4.5}$ power law is shown for reference. The intensity dip seen at 1.8 MeV is an instrumental artifact.

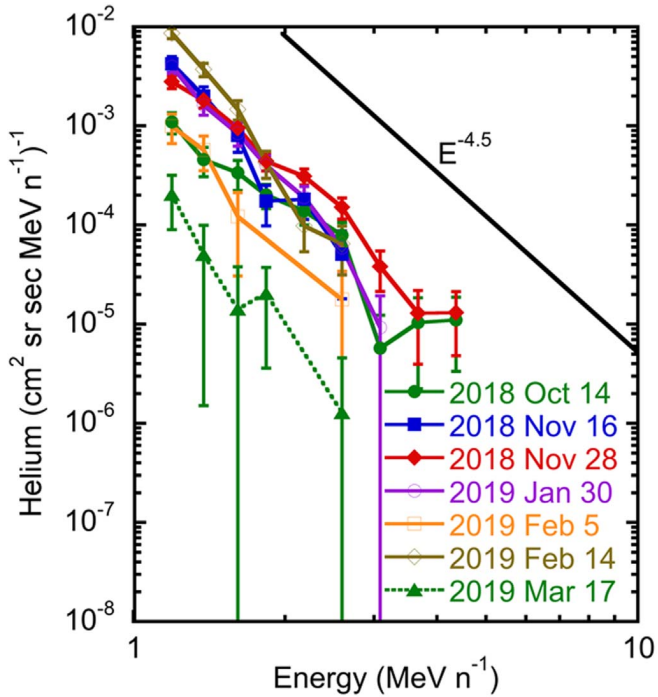

**Figure 5.** Time-integrated helium spectra (from the average of LETA and LETB He spectra) for all seven events. A line corresponding to an $E^{-4.5}$ power law is shown for reference.

*PSP* solar wind plasma and field data (from SWEAP and FIELDS) are only available for events 1-3 and 7. The solar wind structures during two of these time periods, 2018 November 16 and 2018 November 28, are discussed in detail in Allen et al. (2020). The evolution of the solar wind speed during the first three of our events is illustrated in Figure 7. The colored spirals are representative of Parker spiral magnetic field lines corresponding to the measured solar wind speed hourly along the orbit (with the color indicating the solar wind speed according to the color bar on the right side). Superimposed on the spacecraft orbit is the LETA 1–2 MeV proton rate in a logarithmic gray scale. In this representation, it is clear, as in Figure 2, that events 1 and 3 (2018 October 14 and 2018 November 28) occur at approximately the same solar longitude. Both of these events begin in the trailing edge of a fast speed stream, which is followed by a slow speed stream creating a rarefaction region in which the proton rates peak. In contrast, event 2 (2018 November 16) begins near an SIR where a fast stream overtakes a slower stream and peaks within the fast solar wind.

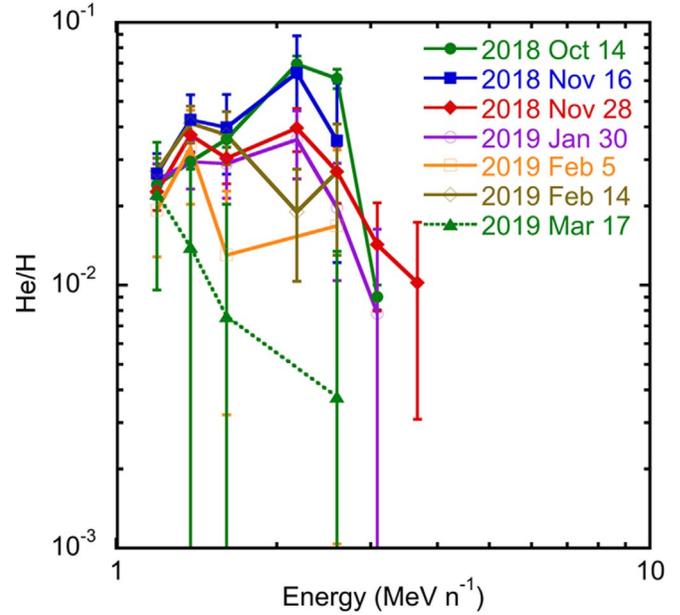

**Figure 6.** He/H abundance ratios vs. energy calculated from the He and H spectra shown in the previous two figures. Given the large uncertainties, it is not possible to determine if there is an energy dependence to the ratio for any of the events.

### 4. Discussion

We have examined the plasma data available for four of our energetic particle events for evidence of stream interactions and compression regions and/or shocks (although the latter would be extremely unusual for a CIR well inside 1 au; see e.g., Van Hollebeke et al. 1978; Jian et al. 2008). Unfortunately, the solar wind data are not available prior to the start of event 1, making it difficult to determine if an SIR preceded the energetic particle event, but SIRs were found prior to or during events 2, 3, and 7; Allen et al. (2020) investigate the plasma properties of the first two of these and the corresponding structures seen by the *Solar Terrestrial Relations Observatory*-A (*STEREO*-A) after further corotation. While increases in the solar wind density and magnetic field are observed, as expected, none of the stream interfaces exhibit a shock. Although Giacalone et al. (2002) show that a fully formed shock is not necessary for particle acceleration, the fact that the energetic particle enhancements do not begin or peak at a clear solar wind structure (as indicated by the in situ plasma and fields measurements) suggests that they are not a result of local/near shock acceleration. More plausibly, the particles are accelerated at larger distances from the Sun where SIRs may have developed into either shock pairs or at least significant compression regions.

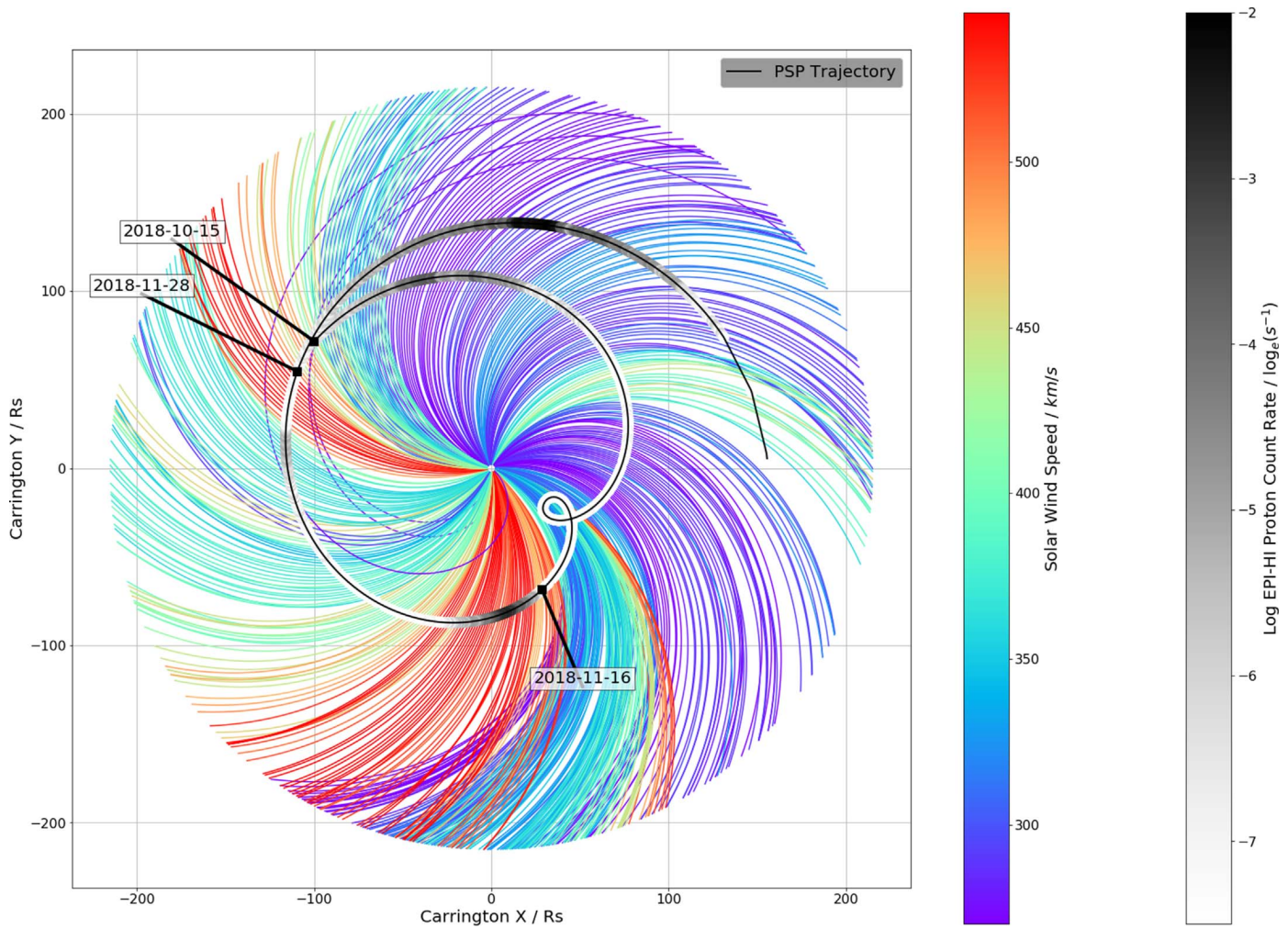

**Figure 7.** Carrington plot of the first *PSP* orbit, with the gray scale (the second color bar on the right) corresponding to the log of the ∼1–2 MeV proton counting rate from LETA. At each hour along the orbit a Parker spiral field line is calculated from the observed solar wind speed from SWEAP and plotted connecting the Sun to *PSP* and beyond with a color corresponding to the solar wind speed (the first color bar on the right). Date labels indicate the start of the first three events of this study (with the exception of the first event where the date label indicates the start of available solar wind data). It should be noted that the Parker spirals for days beyond 2018 November 30 are not drawn and that field lines sampled at the same solar longitudes (e.g., during the orbital corotation loop prior to the 2018 November 16 event) are overplotted and, therefore, overlap at low heliocentric distances. The increased density of field lines in proximity to the loop indicates *PSP*'s slow longitudinal motion during these times.

To gain a larger-scale view of the solar wind structures during each event as well as to compensate for the lack of *PSP* solar wind measurements during events 4–6, Wang–Sheeley–Arge (WSA)-ENLIL simulations (Arge et al. 2004; Odstrcil et al. 2004) were performed for each time period using zero-point corrected quick reduced magnetogram synoptic maps from the Global Oscillation Network Group (GONG). For the events where *PSP* plasma data are available, the simulated *PSP* encounters of slow/fast streams and SIRs match the observations well (Allen et al. 2020). Individual plots of the equatorial plane velocity at the start, midpoint, and end of each event are given in Figure 8. In all of these plots the Sun–*PSP* line is fixed and zebra lines show the magnetic connection between the Sun and *PSP* (as well as to Earth and each *STEREO* spacecraft). Only event 2 (2018 November 16) exhibits what might be expected from a "classic" CIR/SIR event (as suggested by Figure 7). The event starts when *PSP* is magnetically connected along the leading edge of a high-speed stream, continues as the high-speed stream corotates over the spacecraft, and ends close to the trailing edge of the stream. A more complicated variant is seen in event 5 (2019 February 5). Here, the energetic particle intensities start to increase before an initial stream interface passes over *PSP*. The event continues through the first high-speed stream; then through another, stronger stream interface; and ends midway, through a second high-speed stream. There is no obvious "double peak" in the particle intensities that potentially corresponds to this second stream interface.

As mentioned before, there is no evidence of local particle acceleration in either of these events (at least at these energies; see Allen et al. 2020 for a discussion regarding lower energy measurements from EPI-Lo), although it is possible that beyond 1 au, there is a shock(s) associated with the SIR that is capable of accelerating particles. Accelerated particles could then propagate and eventually result in an isotropic distribution that fills a region through which *PSP* then passes. This would be consistent with the observation of similar intensity spectra evident from all three LET apertures, despite their opposing/orthogonal look directions and the lack of velocity dispersion (Figure 2). It could also lead to the erosion of any signature of multiple acceleration sites resulting from the dual SIRs in event 5.

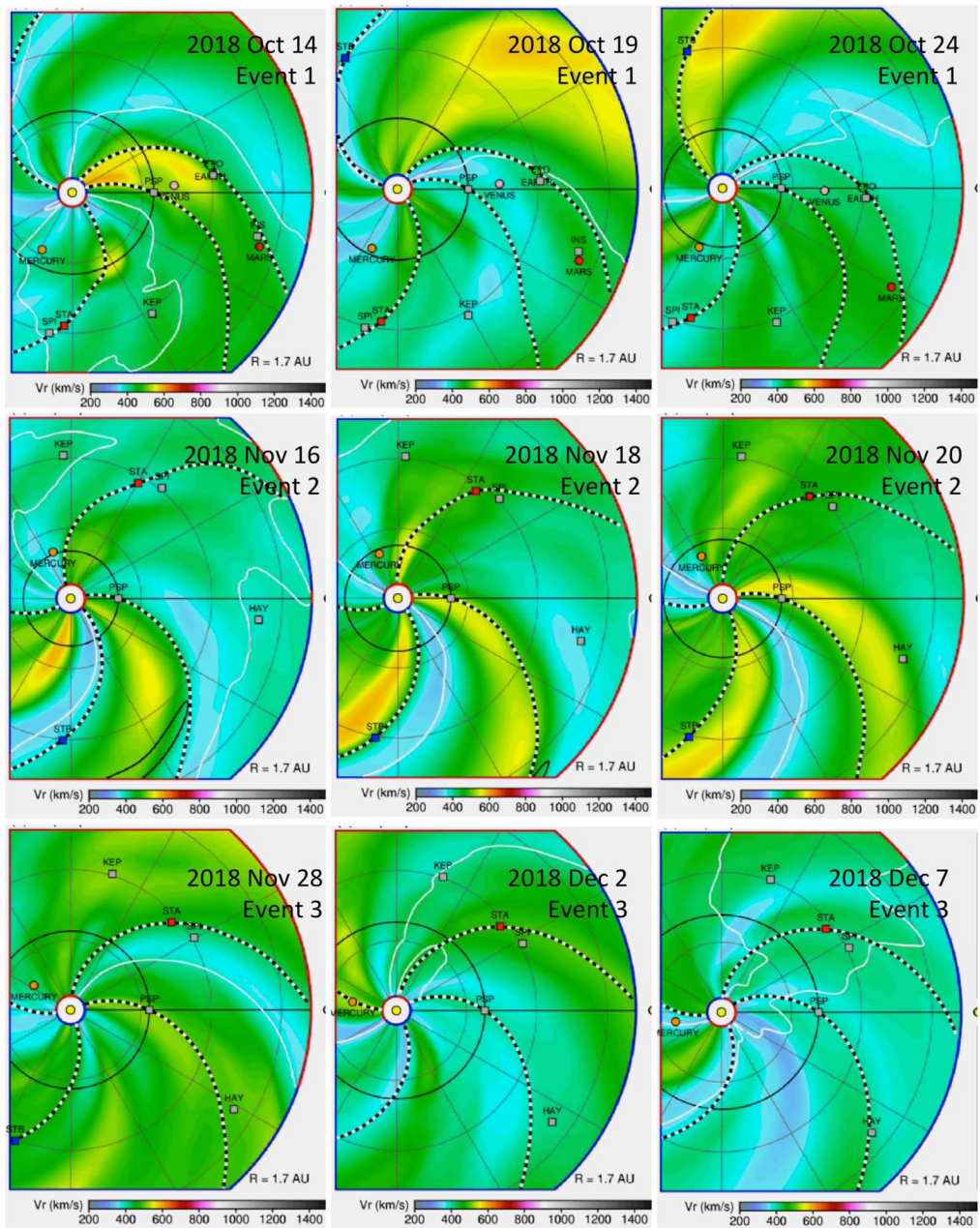

**Figure 8.** Solar wind velocity distributions in the ecliptic plane resulting from WSA-ENLIL simulations at the start (left panels), midpoint (center panels), and the end (right panels) of each energetic particle event. Dates for each snapshot, along with the event number, are given in each panel. The color bar indicates the radial solar wind speed. Zebra lines show the magnetic field connection between the Sun and *PSP* (gray square)/Earth (green circle)/*STEREO*-A (red square)/*STEREO*-B (blue square). The black contours identify CMEs.

Unfortunately, the remaining events are more difficult to interpret in terms of SIRs at *PSP*. Interestingly, they primarily occur during periods of declining or slow solar wind speed, well after the SIR and the majority of the high-speed stream have passed. Additionally, in events 3, 4, and 7 (2018 November 28, 2019 January 30, and 2019 March 17) the SIR appears weak with the high-speed stream of $\sim$200 km s$^{-1}$ faster than the slow wind ahead of it. The majority of events 1 (2018 October 14) and 4 (2019 January 30) occur in the slow wind region between two SIRs; a strong SIR followed by a weak SIR in the case of event 1 and the reverse in the case of event 4. It is possible that these bounding structures are conducive to retaining a region of energetic particles.

In contrast, events 3, 6, and 7 (2018 November 28, 2019 February 14, and 2019 March 17) all end in a region of slow solar wind, well past the initial SIR and associated fast solar wind stream. In a study of CIR-related helium intensities, Reames et al. (1997) found that, at 1 au, the intensities of >1 MeV particles continued to increase well past the high-speed stream, while <1 MeV particle intensities had returned to background levels much earlier. In fact, they observed elevated >1 MeV intensities for more than a week beyond the passage of the SIR and into regions of slow solar wind. This extended presence of the higher energy particles was attributed to the

**Figure 8.** (Continued.)

CIR shock strengthening at larger radial distances where it was capable of accelerating particles to higher energies. Because of their higher rigidities, only the higher energy particles were capable of returning to 1 au in sufficient numbers to register as elevated intensities. It is possible that an analogous situation is responsible for our events that occur in regions of slow solar wind, suggesting acceleration sources at larger radial distances than those of the other events.

All of the events studied here, with perhaps the exception of event 7 (2018 March 17), have softer spectra than the $E^{-4}$, which is typical for CIR events observed at 1 au (Richardson 2004). Whether this is related to the acceleration conditions or the transport conditions is unknown. If the former, then the soft spectra imply weaker shocks than is typical for CIR events observed at 1 au. A transport explanation would suggest less diffusion of the lower energies, relative to the higher, out of the populated region. There does not appear to be any correlation between the spectral index of our events and the solar wind structure in which they occur (per available *PSP* observations and the WSA-ENLIL simulations).

The average He/H abundance ratio quoted by Richardson (2004) for CIR events is 0.058, which is substantially higher than those obtained for any of the events studied here. However, Richardson et al. (1993) found that the He/H abundance varies with the speed of the fast solar wind stream. They obtained an He/H ratio of 0.023 for solar wind streams with a velocity $\leqslant 600$ km s$^{-1}$ and 0.053 for streams with higher velocities. As nearly all the fast solar wind streams in our events have velocities $\leqslant 600$ km s$^{-1}$, the derived abundance ratio values of 0.016–0.031 seem reasonable. It is also possible that the lower He/H abundance is a result of a radial dependence (e.g., Franz et al. 1999); however, our limited sample does not reflect this, consistent with previous studies inside 2.2 au (Christon & Simpson 1979).

Using the WSA-ENLIL simulations as a guide, we have surveyed the energetic particle data available near 1 au in an effort to identify events that are associated with the corotated solar wind structures previously seen on *PSP* in connection with our selected events. For event 1, the Earth is located close to the Sun–*PSP* line, with its magnetic field line close to the leading edge of the high-speed stream that passed *PSP* before the start of the event. Allen et al. (2020) identify a corresponding SIR seen in *Advanced Composition Explorer* (*ACE*) plasma data; however, there is nothing seen in the data from the Ultra Low Energy Isotope Spectrometer (ULEIS; Mason et al. 1998) on *ACE* that can be clearly identified as a CIR particle event during this time. For the remaining events, *STEREO*-A is the most well-positioned spacecraft for comparison. Events 5-7 do not have any particle increases in either the Suprathermal Ion Telescope (SIT; Mason et al. 2008) or the Low Energy Telescope (LET; Mewaldt et al. 2008) that can be convincingly related to our events (there is a particle event seen by both SIT and LET on 2019 February 21; however, this is later than would be expected to correspond to our event 6).

As mentioned previously, Allen et al. (2020) have determined that the stream interfaces of events 2 and 3 arrive at *STEREO*-A on 2018 November 22 and 2018 December 1, respectively. There is an increase of 190 keV nuc$^{-1}$ He intensities observed by SIT on 2018 November 22 lasting for ∼2.5 days, suggesting an event corresponding to our event 2 (2018 November 16). At higher energies, there is no evidence of an increase in the few MeV nuc$^{-1}$ ions observed by LET, probably an indication of a very soft event. In contrast, SIT does not register any particle increases on 2018 December 1, while LET does; however the particle intensity is already elevated due to an onset on 2018 November 29. Whether the additional slight increase on 2018 December 1 is a separate event or merely a temporal variation in the ongoing previous event is unclear, making it difficult to uniquely match the *STEREO*-A event with event 3 (2018 November 28) on *PSP*. Similarly, SIT does not see any particle intensity increase that can be related to event 4 (2019 January 30), while LET registers an increase in the $\geqslant 1$ MeV proton intensities starting around 2019 February 2, with an additional increase on 2019 February 4.

It should be noted that we are currently relying on the WSA-ENLIL simulations for these *PSP*/*STEREO*-A correspondences and, in some cases, the modeled solar wind time profiles do not match those of *STEREO*-A very well (see Allen et al. 2020). A study that utilizes the measured 1 au solar wind more directly to determine the arrival of the SIRs might provide alternate interpretations regarding our events and their potentially corresponding events on *STEREO*-A. The further addition of EPI-Lo observations could provide a more complete picture of the energetic particle spectra and, in combination with *STEREO*-A/SIT+LET spectra, could yield interesting information regarding the radial and/or longitudinal gradient of CIR-related energetic particles. For events 2 and 3, *PSP* was well inside 1 au (Table 1), while it was near 1 au during events 4 and 5 but is separated from *STEREO*-A by ∼60° in the solar longitude. Such a study is beyond the scope of this work but is a possible future investigation.

## 5. Summary

We have detailed the first observations made by *PSP* of energetic particle events associated with SIRs during the first two orbits. Several of these events occurred at locations inside 0.65 au (with two inside 0.5 au), adding to the limited catalog of these events observed well inside Earth's orbit, having been previously only measured by Helios. Two of these events appear to be verified CIR events in that *PSP* observed them on sequential passes of the same solar longitude. All the events are relatively small with measurable proton intensities extending to only a few MeV and have spectra corresponding to power laws with indices ranging from −4.3 to −6.5. This is softer than is typical for CIR events observed near 1 au, but it is not possible to determine whether this is related to the acceleration conditions or to transport effects.

The helium spectra are similarly soft, and we obtain He/H abundance ratios near 1 MeV nuc$^{-1}$, varying somewhat from event to event, with values between $0.016 \pm 0.007$ and $0.031 \pm 0.003$. These ratios are lower than the average He/H abundance of CIR events at 1 au. However, they are consistent with the ratio obtained by Richardson et al. (1993) for CIR events in which the speed of the fast solar wind stream was $\leqslant 600$ km s$^{-1}$. Nearly all the fast solar wind periods observed during our events are also below or close to 600 km s$^{-1}$.

None of the events occurred near a clear solar wind shock/compression region (either in the data or simulations) and all exhibited particle isotropy and a lack of clear velocity dispersion. Thus, we find it likely that these events are not a result of local acceleration but rather correspond to regions previously filled with energetic particles sweeping over the spacecraft. In some cases, it appears that a similar particle population later passes over the *STEREO*-A spacecraft near 1 au, although a detailed study comparing the properties of these populations has yet to be done. Given the currently quiet solar conditions, it is likely that there will be additional events observed during upcoming *PSP* orbits, potentially allowing a more direct and in depth comparison of 1 au observations and those of *PSP*/IS☉IS.

This work was supported by NASA's *Parker Solar Probe* Mission, contract NNN06AA01C. We thank all the scientists and engineers who have worked hard to make *PSP* a successful mission. In particular, we thank B. Kecman and W.R. Cook, without whom the EPI-Hi instrument would not be possible. S.D.B. acknowledges support of the Leverhulme Trust Visiting Professorship program. We gratefully acknowledge the test and calibration support provided by Michigan State University's National Superconducting Cyclotron Laboratory, Texas A&M University's Cyclotron Institute, and the Lawrence Berkeley National Laboratory's 88-inch Cyclotron Laboratory.

## ORCID iDs

E. R. Christian ● https://orcid.org/0000-0003-2134-3937
A. C. Cummings ● https://orcid.org/0000-0002-3840-7696
M. I. Desai ● https://orcid.org/0000-0002-7318-6008
M. E. Hill ● https://orcid.org/0000-0002-5674-4936
C. J. Joyce ● https://orcid.org/0000-0002-3841-5020
R. A. Leske ● https://orcid.org/0000-0002-0156-2414
W. H. Matthaeus ● https://orcid.org/0000-0001-7224-6024
D. J. McComas ● https://orcid.org/0000-0001-6160-1158
D. G. Mitchell ● https://orcid.org/0000-0003-1960-2119
J. S. Rankin ● https://orcid.org/0000-0002-8111-1444
N. A. Schwadron ● https://orcid.org/0000-0002-3737-9283
J. R. Szalay ● https://orcid.org/0000-0003-2685-9801
R. C. Allen ● https://orcid.org/0000-0003-2079-5683
L. K. Jian ● https://orcid.org/0000-0002-6849-5527
D. Lario ● https://orcid.org/0000-0002-3176-8704
S. D. Bale ● https://orcid.org/0000-0002-1989-3596
S. T. Badman ● https://orcid.org/0000-0002-6145-436X
M. Pulupa ● https://orcid.org/0000-0002-1573-7457
R. J. MacDowall ● https://orcid.org/0000-0003-3112-4201
J. C. Kasper ● https://orcid.org/0000-0002-7077-930X
A. W. Case ● https://orcid.org/0000-0002-3520-4041
K. E. Korreck ● https://orcid.org/0000-0001-6095-2490
M. L. Stevens ● https://orcid.org/0000-0002-7728-0085